\documentclass[10pt,letterpaper,twocolumn]{article} %% two column, final layout
\usepackage{ol2}
\usepackage[draft,implicit=false]{hyperref}
\usepackage{amsmath}
\usepackage{hyperref}

\begin{document}

\twocolumn[ %% activate for two-column option

\title{Direction-sensitive transverse velocity measurement by phase-modulated structured light beams}

\author{Carmelo Rosales-Guzm\'an,$^{1,*}$ Nathaniel Hermosa,$^1$ Aniceto Belmonte$^2$ and Juan P. Torres$^{1,2}$}

\address{
 $^1$ICFO-Institut de Ciencies Fotoniques, Mediterranean Technology Park, 08860 Castelldefels (Barcelona), Spain\\
 $^2$Universitat Polit\`{e}cnica de Catalunya, BarcelonaTech, Dept. of Signal Theory \& Communications, 08034 Barcelona, Spain \\
$^*$Corresponding author: \href{mailto:carmelo.rosales@icfo.es}{carmelo.rosales@icfo.es}
}

\begin{abstract}
The use of structured light beams to detect the velocity of targets moving perpendicularly to the beam's propagation axis opens new avenues for remote sensing of moving objects. However, determining the direction of motion is still a challenge since detection is usually done by means of an interferometric setup which only provides an absolute value of the frequency shift. Here, we put forward a novel method that addresses this issue. It uses dynamic control of the phase in the transverse plane of the structured light beam so that the direction of the particles' movement can be deduced. This is done by noting the change in the magnitude of the frequency shift as the transverse phase of the structured light is moved appropriately. We demonstrate our method with rotating micro-particles that are illuminated by a Laguerre-Gaussian beam with a rotating phase about its propagation axis. Our method, which only requires a dynamically configurable optical beam generator, can easily be used with other types of motion 
by appropriate engineering and dynamic modulation of the phase of the light beam.
\end{abstract}

] %% activate for two-column option

%\section{Introduction}
The Doppler effect is the basis of a myriad of laser remote sensing systems aimed at monitoring the velocity of moving targets \cite{LaserRemote}. In a standard monostatic laser remote system where the transmitter and the receiver are at the same location, the classical longitudinal Doppler effect only allows detection of the velocity of the target along the line-of-sight between the emitter and the target ({\em longitudinal velocity}); any transverse velocity component generates no frequency shift.

Even though this is not a major concern in many applications, in some circumstances however, this can severely limit the use of remote sensing systems based on the Doppler effect. One can resort to perform many rapid Doppler measurements along a large set of different directions, effectively transforming undetectable transverse velocities to have a longitudinal component which can thus be detected \cite{ApplOpt1982}. In general, implementing these schemes are cumbersome because they require fast mechanical realignment of the direction of propagation of the laser beam in a time scale that should be shorter than the movement of the target.

There exists a relativistic transverse Doppler effect which is sensitive to the transverse and longitudinal velocities \cite{Sommerfeld}. However, it yields frequency shifts that are staggeringly small ($\sim v^2/c^2$) for typical velocities involved in most laser remote sensing applications of interest ($v \ll c$, where $v$ the velocity of the target and $c$ the velocity of light).

Recently, it has been demonstrated \cite{ol,Science,SciRep} that targets moving in a plane perpendicular to the direction of illumination by the laser beam can induce a velocity-dependent frequency shift if the illuminating beam is a structured light beam, \textit{i.e.}, it contains an appropriate transverse phase profile. This frequency shift, which is on top of the usual longitudinal Doppler shift, enables determination of the velocity component perpendicular to the illumination axis ({\em transverse velocity}).

The value of the generated frequency shift depends on the velocity of the target and on the specific phase profile imprinted on the light beam.  For the case of a purely rotational motion with angular velocity $\Omega_t$, the appropriate phase gradient of the illuminating beam should preferably have circular symmetry, such as in the case of a Laguerre-Gauss beam,
\begin{equation}
E(\rho,\varphi,t)=U(\rho)\exp(ikz+il\varphi-i\omega t) + h.c.
\end{equation}
where $z$ is the direction of propagation of the light beam, $\omega=ck$ is the angular frequency, $k$ is the wavenumber, $\rho$ and $\varphi$ are the radius and the azimuthal angle in cylindrical coordinates, respectively, $U(\rho)$ is the radial profile and the winding number $\ell$ is the number of times the phase jumps from $0$ to $2\pi$ as one goes around the azimuthal angle $\varphi$.

In this case, the frequency shift $\Delta f$ is given by
\cite{ol},
\begin{equation}
\Delta f= \frac{\ell\Omega_t}{2\pi}. \label{Rotation}
\end{equation}
The validity of this expression has been experimentally demonstrated in two recent experiments \cite{Science,SciRep}. In \cite{Science}, the target was illuminated with two co-propagating beams with opposite winding number, $+\ell$ and $-\ell$, and the frequency of the beating of the two beams is measured. In \cite{SciRep}, $\Delta f$ is determined through an interferometric technique where light reflected from the moving target is made to interfere with a reference signal.

Since most photodetectors are sensitive to light intensity alone and the detection systems are interferometric in nature, only the absolute value of the difference between the frequency of the reflected and the reference light can be obtained, \textit{i.e.} there is no indication of whether the received scattered wave is upshifted or downshifted in frequency with respect to the frequency of the incident beam. Therefore, information about the direction of motion, clockwise or anti-clockwise in the case of rotational motion, is not available.

Well known techniques for direction-sensitive velocity measurements are generally based on the generation of an optical frequency offset between the illumination and the reference beam ({\em heterodyne detection}) or between the two illumination beams. For this purpose, the use of mechanically rotating waveplates \cite{JOSA} or diffraction gratings \cite{ApplOpt} have been extensively reported. Other techniques employ acousto-optic modulators\cite{ElectronLett} or electro-optic frequency shifters \cite{AppOpt1978}. In general, these systems should be customized for a particular beam size and specific frequency.

\begin{figure}[t]
\centering
\includegraphics[width=.48\textwidth]{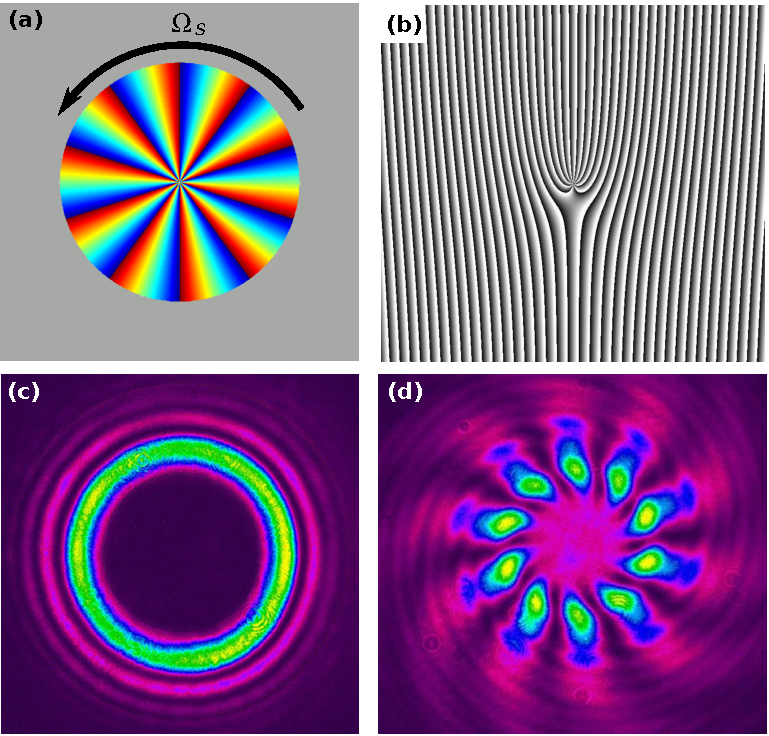}
\caption{\small (Color online). (a) The illuminating beam's phase is rotated either clockwise or anti-clockwise with angular velocity $\Omega_s$ as illustrated here. The phase changes from zero (blue) to $2\pi$ (red) ten times around the azimuth. (b) Fork-like hologram displayed in the SLM to generate the $LG^{10}_0$ mode. (c) Experimental intensity profile of a Laguerre-Gaussian $LG^{\ell}_0$ beam with winding number $\ell=10$, used to illuminate the moving targets. (d) Interference pattern between the $LG^{10}_0$ and a Gaussian beam obtained in experiments.} \label{Beam}
\end{figure}

The aim of this paper is to put forward a novel method to detect transverse velocities that can discriminate velocity direction based on the use of properly modulated phase in the transverse plane of a structured light beam. In a sense, the scheme described here plays a similar role in the transverse plane to the role of an acousto-optic modulator, which dynamically change the phase along the propagation direction of the source light beam ({\em longitudinal phase}) modifying its frequency. In our method, the phase change takes place in the transverse plane of the illuminating beam. The significance of this method lies in the fact that: 1) absolute velocity directions can be determined easily; and 2) it does not require the use of additional components other than a dynamic and configurable optical beam generator. Here, we applied our method to a specific case: rotation of micro-particles. This technique can easily be generalized to other types of motion by proper tailoring and controlling of the movement of 
the phase of the illumination beam.

\begin{figure}[t]
\centering
\includegraphics[width=.48\textwidth]{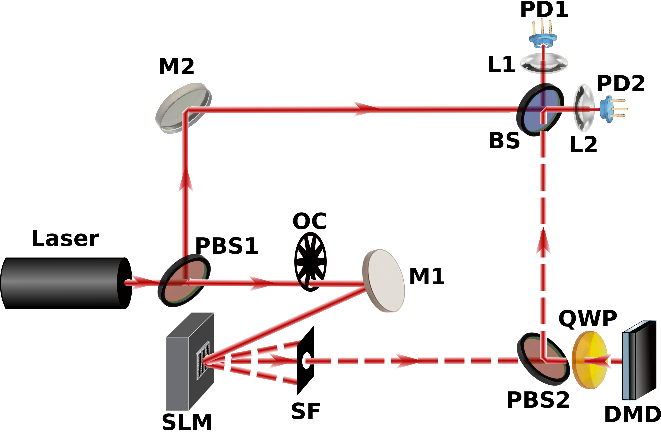}
\caption{\small Experimental setup to extract the rotation velocity and its sense of direction. PBS: polarizing beam splitter; M: mirror; L: lens; PD: photodetector; SLM: Spatial Light Modulator; SF: spatial filter; OC: optical chopper; QWP: Quarter-Wave  Plate; DMD: Digital Micromirror Device. {\it See text for details.}}
\label{setup}
\end{figure}
%\section{Experimental setup}

For example, in the special case of a rotating target, a rotating helical-phased beam can be used to extract information about its rotation direction. The phase of the beam now takes the form $\Phi(t)=\ell \varphi \pm \ell \Omega_s t$, where $\Omega_s$ is the velocity of rotation of the phase [see Fig. \ref{Beam}(a)]. The time-varying light beam can be obtained by sequentially displaying holograms calculated for different rotated phase. A sample of the fork-like hologram displayed in the Spatial Light Modulator is shown in Fig. \ref{Beam}(b). Figures \ref{Beam}(c) and (d) show the experimental intensity and phase profile of the Laguerre-Gaussian mode with winding number $\ell=10$. The phase is obtained after interference with a Gaussian beam. Since $\ell=10$, the phase jumps from $0$ to $2\pi$ ten times as one goes from $\varphi=0$ to $\varphi=2\pi$. The frequency shift of the light reflected back from the target is now given by,
\begin {equation}
\Delta f^{\prime}=\frac{\ell(\Omega_t - \Omega_s)}{2\pi}.
\label{Relative}
\end{equation}
One has $|\Delta f^{\prime}| < |\Delta f|$ when both $\Omega_t$ and $\Omega_s$ has the same sign, whereas $|\Delta f^{\prime}| >|\Delta f|$ when they have opposite signs. A higher frequency shift will be measured when the target rotates in the opposite direction of the beam.

We extract the Doppler frequency shift imparted by the moving particles via an interferometric technique using the modified Mach-Zehnder interferometer shown in Fig. \ref{setup}. A $15$ mW continuous wave He-Ne laser (Melles-Griot, $\lambda=632.8$ nm) is spatially cleaned and expanded to a diameter of $5$ mm, using a lens combination of focal lengths $f=25$mm and $f=100$mm for the front and back lens, respectively, and a $30\mu$m pinhole placed at the middle focus. This beam is split into two ({\em signal} and {\em reference} beams) using a polarizing beam splitter (PBS1). A mirror (M1) redirects the signal beam to a Spatial Light Modulator (SLM, LCOS-SLM X10468-02 from Hammamatsu) that imprints the beam with the desired phase profile, as shown in Fig. \ref{Beam}(a). The first diffracted order of a fork-like hologram [Fig. \ref{Beam}(b)] encoded into the SLM is used to illuminate the target while the rest are spatially filtered (SF) using two lenses of focal lengths $f=50$mm and a $200\mu$m pinhole 
placed at the middle focus. A second polarizing beam splitter (PBS2) in combination with a quarter-wave plate (QWP) collects light reflected from the target back into the interferometer. These reflections are afterwards interfered with the reference signal using a beam splitter (BS).

A balanced detection is implemented with two photodetectors (PD1 and PD2). These are connected to an oscilloscope (TDS2012 from Tektronix). An optical chopper (OC) placed in the path of the signal beam shifts our detected frequency from Hz to kHz, so that we can eliminate low frequency noise. An autocorrelation process that cross-correlates the signal with itself to find periodic patterns obscured by noise, allows us to enhance the signal-to-noise ratio. The resulting signal is Fourier transformed to find the frequency content. Hamming windowing and zero padding are also applied to smooth the Fourier spectrum.

\begin{figure}[t]
 \centering
\includegraphics[width=.35\textwidth]{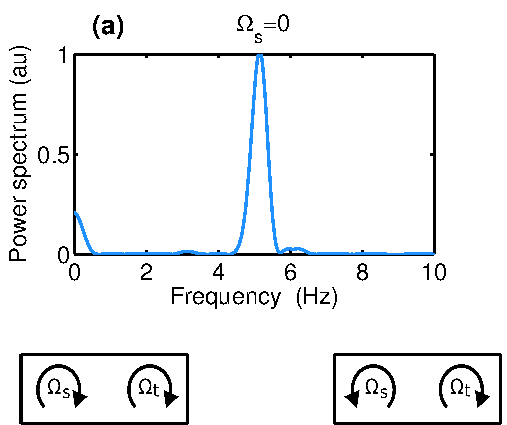}
\includegraphics[width=.48\textwidth]{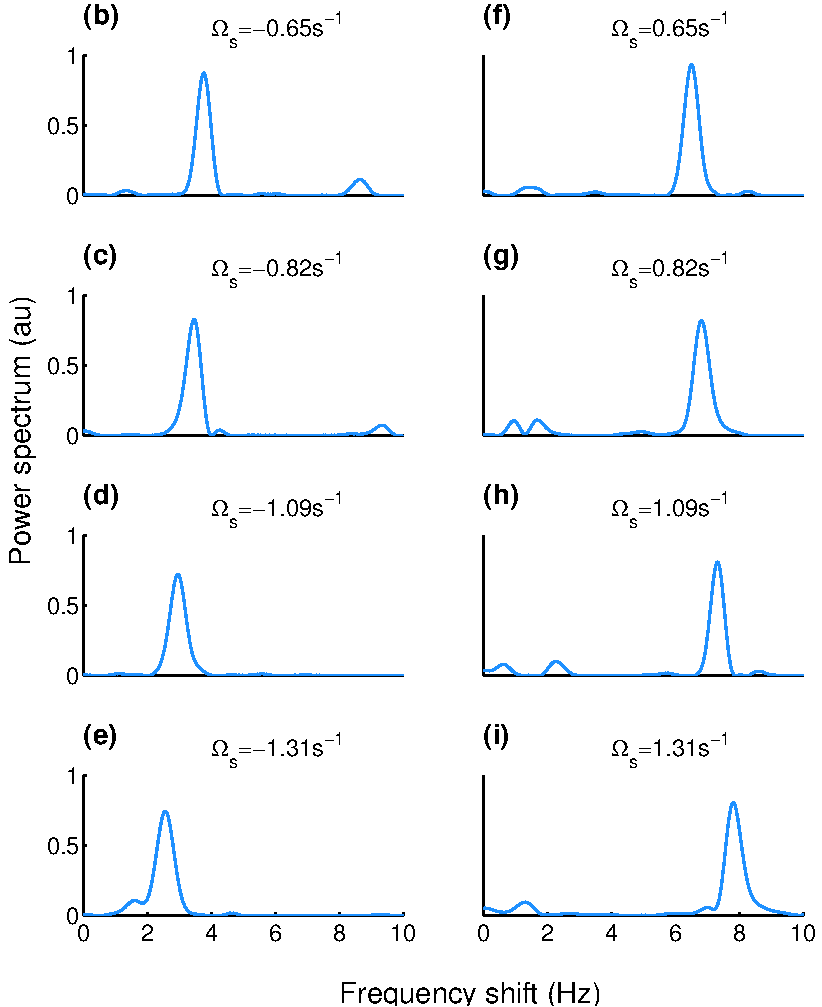}
 \caption{\small Fourier spectrum of the detected signal. In (a) the phase of the $LG^{10}_0$ is static. From (b) to (e) the phase of the $LG^{10}_0$ is rotated anti-clockwise with increasing angular velocities. From (f) to (i) it is rotated clockwise, also with increasing angular velocities.}
\label{spectrum}
 \end{figure}

The rotation of the particles was mimicked using a Digital Micromirror Device (DMD) as in \cite{SciRep}. For this experiment we used an arrange of 7x14 micromirrors to simulate a disk-like particle of 70 $\mu m$ in diameter. Its rotation was produced by displaying consecutive binary images of the  particle at a position separated from the previous by an amount $\Delta \theta=2\pi/N$,  N is the number of images to complete a $2\pi$ rotation, which in our case was 96. Hence, the angular velocity will be $\Omega_t=\Delta\theta/\Delta t =2\pi/(NT)$, T is the time interval between consecutive images.

Different methods on how to rotate a beam have been reported \cite{PhysRevLett,OptExpress2008,OptCommmun ,JOptAPureApplOpt}. Here we programmed the SLM to display consecutive images of the fork-like hologram that produces the $LG^{10}_0$ mode. In each image, the phase appears rotated by an amount $\Delta \varphi=2\pi/\eta$, where $\eta$ is an integer number. The time from one image to the next is the refresh rate $\tau$ of the SLM. The angular velocity of rotation of the phase can then be computed as $\Omega_s=2\pi/(\eta\tau)$. In order to increase (decrease) the angular velocity $\Omega_s$, we can increase (decrease) $\eta$ or $\tau$. In any case, $\eta$ should be larger than $2\ell$ to avoid aliasing of the generated signal. Even though most commercially available SLMs have limited frame refresh rate of 60 Hz, with the emerging technology of DMDs, refresh rates of up to 4 KHz can be achieved \cite{OptExpress,OptExpress2006}.

\begin{figure}[t]
\centering
\includegraphics[width=.47\textwidth]{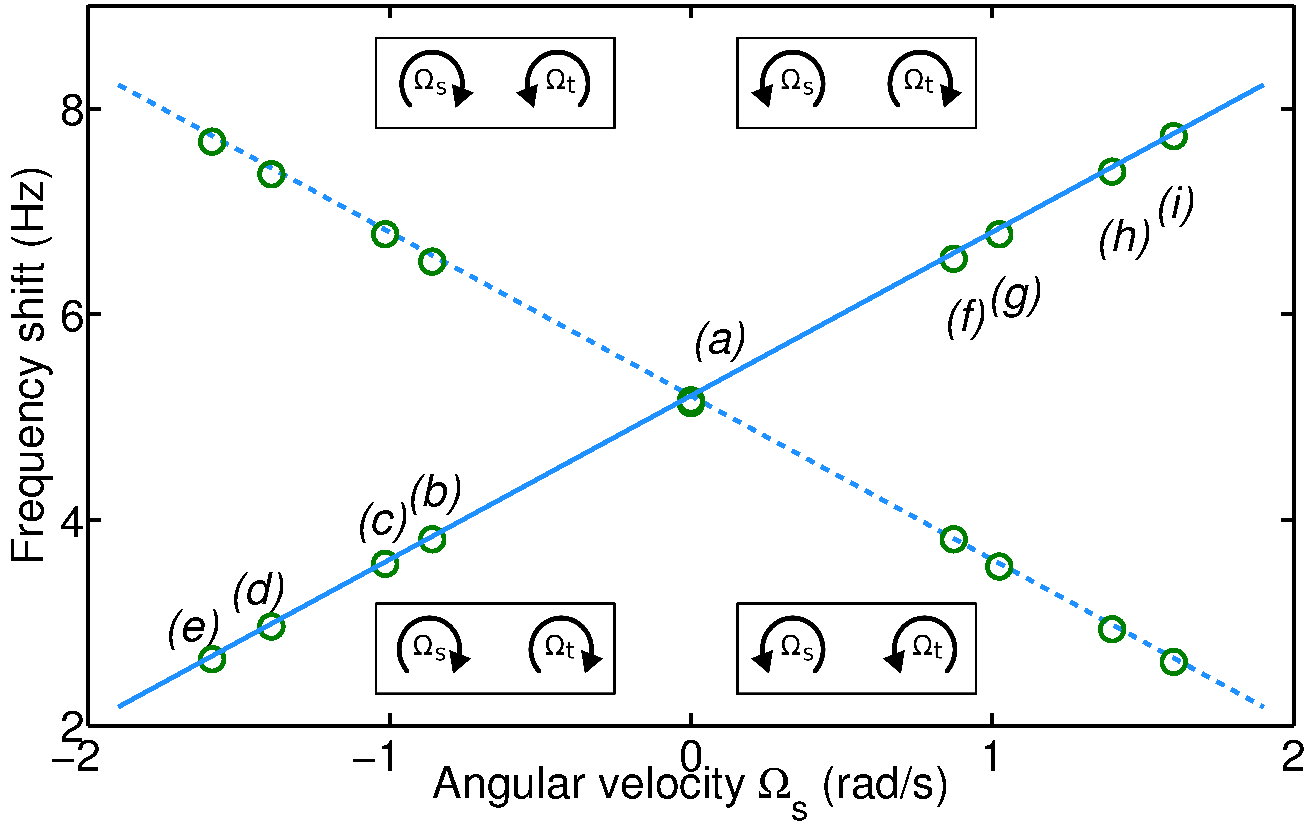}
\caption{\small Frequency shift as function of the angular velocity of the phase gradient of the illuminating beam. The particle rotates clockwise (solid line) or anti-clockwise (dashed line). Labels (a) to (i) corresponds to the same cases depicted in Fig. \ref{spectrum}}
 \label{four}
\end{figure}

Figure 3 shows the spectra produced with a $LG_0^{10}$ beam with a static phase, and a $LG_0^{10}$ beam with rotating phase at different angular speeds. In Fig. \ref{spectrum}(a), the frequency shift is induced only by the movement of the particles as expected. The frequency obtained is $\Delta f= 5.2 s^{-1}$. This frequency decreases when the phase is rotated in the same direction as the rotation sense of the particle [Fig.\ref{spectrum}(b)-(e)] while it increases when the phase is rotated in the opposite direction [Fig.\ref{spectrum}(f)-(i)]. Moreover, the amount of frequency shift with respect to a non-rotating phase is determined by the rate of the rotation of the $LG^{10}_0$ phase. Direction sensitivity is clear as the frequency is upshifted or downshifted based on the relative direction of rotation of the $LG^{10}_0$ phase.

In Figure \ref{four}, we plot the frequency shift $\Delta f^{\prime}$ as function of the velocity of rotation of the phase ($\Omega_s$). We do this for two possible directions of rotation, clockwise ($\Omega_t<0$) and anti-clockwise ($\Omega_t>0$). According to Eq. \ref{Relative} there is a linear relationship between $\Delta f^{\prime}$ and $\Omega_s$. This is clearly observed in Fig. \ref{four} for two cases: $\Omega_t> 0$ and $\Omega_t< 0$. As in Fig. \ref{spectrum}, the point labeled $(a)$ corresponds to a static phase ($\Omega_s=0$), whose corresponding frequency shift is $\Delta f= 5.2 s^{-1}$. From $(b)$ to $(e)$ the phase rotates anti-clockwise ($\Omega_s<0$) with increasing angular velocities, whereas from $(f)$ to $(i)$ it rotates in the opposite direction $\Omega_s>0$, again with increasing angular velocities. Whenever $\Omega_t$ and $\Omega_s$ have opposite signs, the generated frequency shift $\Delta f^{\prime}$ will be larger than $\Delta f$. Conversely, if they have the same sign, the 
frequency shift will be smaller.

Figure \ref{four} can be considered a description of the measurement method. For a given value of the sought-after $\Omega_t$, changing the velocity $\Omega_s$ in a controlled way with a programmable spatial phase modulator allows detection of the frequency shift as a function of $\Omega_s$. The dashed line (negative slope) corresponds to a particle with anti-clockwise rotation ($\Omega_t>0$), while the continues line (positive slope) corresponds to a particle rotating clockwise ($\Omega_t<0$).

To summarize, we have presented the experimental demonstration of a direction-sensitive velocity measurement method that uses frequency shifts induced by dynamic structured light beam illumination. In addition to the frequency shifts induced by the moving target under investigation, the frequency is upshifted or downshifted based on the relative direction between the target's movement and the movement of the phase. This enables the detection of the absolute direction of the target's movement. The method is easy to implement since it does not require the use of additional optical elements. It uses the same device that generates the structured light beam. In particular, we implemented the method by obtaining the rotational velocity, its absolute value and its sign, of a rotating particle employing optical beams with a rotating helical phase. Even though we restricted our attention to rotating targets, this method can be easily generalized to more complex movements by properly tailoring the structured 
illumination beam.

C. Rosales-Guzm\'an would like to thank V. Rodr\'iguez-Fajardo for insightful discussions. This work was supported by the Government of Spain (project FIS2010-14831 and program SEVERO OCHOA), and the Fundacio Privada Cellex Barcelona.

%\bibliographystyle{osajnl}
%\bibliography{Bibliography}

\end{document}